\documentclass[twocolumn,prb,showpacs,floatfix]{revtex4}

\usepackage{graphicx}
\usepackage{dcolumn}
\usepackage{bm}
\usepackage{amsmath}

\begin{document}

\title{Edge and bulk components of lowest-Landau-level orbitals, correlated fractional 
quantum Hall effect incompressible states, and insulating behavior in finite graphene samples}

\author{Constantine Yannouleas}
\email{Constantine.Yannouleas@physics.gatech.edu}
\author{Igor Romanovsky}
\email{Igor.Romanovsky@gatech.edu}
\author{Uzi Landman}
\email{Uzi.Landman@physics.gatech.edu}

\affiliation{School of Physics, Georgia Institute of Technology,
             Atlanta, Georgia 30332-0430}

\date{19 July 2010; Phys. Rev. B {\bf 82}, 125419 (2010)}

\begin{abstract}
Many-body calculations of the total energy of interacting Dirac electrons in finite
graphene samples exhibit {\it joint\/} occurrence of cusps at angular momenta
corresponding to fractional fillings characteristic of formation of incompressible
(gapped) correlated states ($\nu=1/3$ in particular) and opening of an insulating energy 
gap (that increases with the magnetic field) at the Dirac point, in correspondence with
experiments. Single-particle basis functions obeying the zigzag boundary condition at the
sample edge are employed in exact diagonalization of the interelectron Coulomb interaction, 
showing, at all sizes, mixed equal-weight bulk and edge
components. The consequent depletion of the bulk electron density attenuates the
fractional-quantum-Hall-effect excitation energies and the edge charge accumulation 
results in a gap in the many-body spectrum.
\end{abstract}


\maketitle

\section{Introduction}

The isolation of graphene sheets \cite{geim04} was soon followed by experiments 
\cite{geim05,kim05} on the anomalous integer quantum Hall effect (IQHE), 
demonstrating the formation of Landau levels, described with non-interacting electron theory. 
Subsequently, theoretical explorations \cite{jain06,chak06} appeared 
pertaining to the fractional quantum Hall effect (FQHE) in graphene, which, in contrast to the
IQHE, is known to be the manifestation of collective correlated interacting electrons 
phenomena. For Dirac electrons in the lowest ($n$ = 0) Landau level (LLL), it was found 
\cite{jain06,chak06} (in a spherical geometry) that the behavior of the incompressible states 
in boundless graphene models emulates that of the well-known FQHE in the semiconductor 2D 
electron gas. \cite{laug83,yann07,jainbook} Recent experiments, \cite{andr09,kim09}
employing small (1 $\mu$m) suspended single-sheet samples have shown that the appearance of 
the FQHE in graphene is accompanied by the emergence of an insulating phase. The physics that 
underlies this joint occurrence, coupled with the systematic theoretical overestimation 
\cite{jain06,chak06} of the measured FQHE excitation energies, remains unresolved; similar 
findings in recent experiments on bilayer graphene \cite{bao10} further highlight these open 
issues. 

In this paper, we show, using exact diagonalization \cite{yann07} (EXD) 
(in the quantum Hall regime) of the many-body hamiltonian
for interacting Dirac electrons occupying the LLL of finite planar graphene samples, 
that the incompressible correlated electron physics is modified [using zigzag boundary 
conditions \cite{geim09} (ZBC)] in two main ways: (i) the FQHE excitation energies are 
significantly attenuated due to depletion of the density of the bulk 
component of the Dirac electrons spinor, and (ii) the concurrent accumulation of charge at the
sample edge underlies the opening of a gap in the many-electron spectrum (increasing with the 
magnetic field, as observed experimentally), relating to the aforementioned insulating phase.

Underlying the above many-body behavior is the existence of LLL single-particle states of 
mixed character, exhibiting bulk- [Darwin-Fock-type \cite{yann07} (DF)] and edge-type 
components of {\it equal weights\/} \cite{note56} ({\it independent of sample size\/}), each 
residing on a different graphene sublattice. With interelectron repulsion, the bulk component 
exhibits FQHE characteristics (with excitation energy gaps reduced by a factor of four compared
to boundless graphene). The edge component is associated with formation of a single-ring 
rotating Wigner molecule \cite{yann07} (RWM). The size-independent influence of the graphene 
edge on the many-body properties ushers a new paradigm, contrasting the commonly accepted 
expectation that materials bulk properties are unaffected by the detailed configuration or 
conditions at the surface. \cite{peierls} This unique behavior derives from the two-sublattice 
topology of the graphene net modeled here by the continuous Dirac-Weyl (DW) equation.
\cite{geim09}

An important element in our approach is formulation of the solutions of the Dirac-Weyl
equation (with the zigzag boundary conditions) using the Kummer confluent hypergeometric
function. \cite{abrabook} This formulation provides systematic analytical insights into the
nature of the DW single-particle states (bulk and edge), as well as it permits efficient and
accurate numerical evaluations (with the help of algebraic computer languages \cite{mathbook}).

In Sec. \ref{secLLLman} we describe the confluent-hypergeometric-function formulation of the 
solutions of the DW equation and construct the LLL DW spinor in terms of equal-weight bulk and 
edge components. Details of the solutions of the DW equation in polar coordinates are given
in Appendix A. 

A brief description of the exact-diagonalization many-body method, and the contributions of
the bulk and edge spinor components to the Coulomb interaction matrix elements are given in
Sec. \ref{secexd}. 

In Sec. \ref{secres} we display results for the ground-state and excited spectra, and their
relation to the $\nu=1/3$ FQHE incompressible state and to the insulating behavior at 
$\nu=0$. 

A summary is presented in Sec. \ref{secsum}.

Finally, in Appendix B we distinguish the mixed bulk-edge Dirac spinor states (forming the 
quasidegenerate LLL manifold in finite graphene samples with ZBC), where one of the spinor 
components is located in the bulk of the sample and the other localized at the edge, from
the ``double-edge'' states where both components of the spinor represent edge states. In the 
``double-edge'' case, one component corresponds to a bulk orbital which transforms (when the 
orbital's centroid falls near the graphene sample boundary) into an edge state familiar from the
theory\cite{halp82} of the nonrelativistic integer QHE. The other component of the 
``double-edge'' spinor, as well as the edge orbital of the aforementioned mixed bulk-edge one, 
corresponds to an edge state unique to the two sublattice topology of graphene. We discuss
and illustrate the natural way in which the above classification of Dirac-electron edge states 
arises within the framework of the Kummer function formalism for the solution of the DW equation
described in Sec. \ref{secLLLman} and Appendix A.

\section{The single-particle lowest-Landau-level manifold with zigzag boundary conditions}  
\label{secLLLman}

We first discuss the solution of the Dirac-Weyl equation in polar coordinates under 
the imposition of the zigzag boundary condition. We model the low-energy noninteracting 
graphene electrons (around a given $K$ point) via the continuous DW 
equation. \cite{geim09} Circular symmetry leads to conservation of 
the total pseudospin \cite{geim09} $\hat{\jmath}=\hat{l}+\hat{\sigma}_z$, where $\hat{l}$ is 
the angular momentum of a Dirac electron. As a result, we seek solutions for 
the two components $\psi^A ({\bf r})$ and $\psi^B ({\bf r})$ (associated with the two 
graphene sublattices $A$ and $B$) of the single-particle electron orbital (a spinor) 
that have the following general form in polar coordinates:
\begin{equation}
\psi_l({\bf r})=
\left( 
\begin{array}{c}
\psi^A({\bf r})\\
\psi^B({\bf r})
\end{array}
\right)=
\left(
\begin{array}{c}
e^{il\phi} \chi^A(r)\\
i e^{i(l+1)\phi} \chi^B(r)
\end{array}
\right).
\label{sp}
\end{equation} 
The angular momentum takes integer values; for
simplicity in Eq.\ (\ref{sp}) and in the following, the subscript $l$ is omitted in 
the sublattice components $\psi^A$, $\psi^B$ and $\chi^A$, $\chi^B$.      
 
With Eq.\ (\ref{sp}) and a constant magnetic field $B$ (symmetric gauge), the DW equation 
reduces (for the $K$ valley) to
\begin{eqnarray}
\frac{d}{dx} \chi^B + \frac{1}{x} \left( l+1 + \frac{x^2}{2} \right) \chi^B
&=& \varepsilon \chi^A \nonumber \\
\frac{d}{dx} \chi^A - \frac{1}{x} \left( l + \frac{x^2}{2} \right) \chi^A
&=& - \varepsilon \chi^B, 
\label{dweq}
\end{eqnarray}  
where the reduced radial coordinate $x=r/l_B$ with $l_B=\sqrt{\hbar c / (eB)}$ the
magnetic length. The reduced single-particle eigenenergies $\varepsilon=E/(\hbar v_F/l_B)$, 
with $v_F$ the Fermi velocity. Since the properties of the solutions of the DW equation with a
magnetic field using the ZBC are not widely known, we outline pertinent details on this
subject in the following (see also Appendices A and B).

For a finite circular graphene sample of radius $R$, we seek solutions of Eq.\ (\ref{dweq}) 
for $\varepsilon \neq 0$ with $l \leq -1$ (corresponding to $j < -1/2$) that are {\it 
regular at the origin\/} ($x=0$). The general form of solution is 
\begin{eqnarray}
\chi^A(x) = - {\cal C}  \frac{\varepsilon}{2|l|} e^{-x^2/4} x^{|l|} 
M\left( 1-\frac{\varepsilon^2}{2}, |l|+1, \frac{x^2}{2} \right),
\nonumber \\
\chi^B(x) = {\cal C} 
e^{-x^2/4} x^{|l|-1} M\left( -\frac{\varepsilon^2}{2}, |l|, \frac{x^2}{2} \right),
\label{solneg}
\end{eqnarray} 
where $M(a,b,z)$ is Kummer's confluent hypergeometric function \cite{abrabook} and ${\cal C}$
is a normalization constant. 
  
The ZBC requires that one sublattice component vanishes at the physical
edge situated at the finite radius $R$ of the graphene sample. Therefore at least one 
of the Kummer functions in Eq.\ (\ref{solneg}) must exhibit zeros (nodes) for $x>0$. The
search for such zeros is greatly facilitated with the use of the following theorem:
\cite{gatt08} For the Kummer function $M(a,b,z)$ ($z$ being real), it is known that, if 
$b > 0$ [which is the case for both Kummer functions in Eq.\ (\ref{solneg})], there are no 
zeros of $M(a,b,z)$ for $z >0$ if $a \geq 0$, and there are precisely $- \lfloor a \rfloor$ 
zeros if $a < 0$; the floor function is defined as: $\lfloor a \rfloor = n$ if and only if 
$n \leq a < n+1$, where $n$ is any positive or negative integer (including zero).

The $\chi^B$ LLL component does exhibit precisely one zero, enabling a transcendental equation 
(for a given $x_R$) for the single-particle energies $\varepsilon$
\begin{equation}
\chi^B(\varepsilon, l, x_R)=0,\;\;\; \text{or} \;\;\;
M\left( -\frac{\varepsilon^2}{2}, |l|, \frac{x_R^2}{2} \right)=0.
\label{eigeq}
\end{equation}

The solutions to Eq.\ (\ref{eigeq}) (for a range of $x_R >> 1$, i.e., high magnetic fields,
and for any $l \leq -1$) give states (which compose the LLL manifold) with near vanishing
energies $0 < \varepsilon/\sqrt{2} << 1$. Consequently, within the graphene sample, when 
$R/l_B >> 1$, $\chi^B$ can be approximated by a bulk-type DF orbital, i.e.,
\begin{equation}
|b\rangle={\cal C} e^{-x^2/4} x^{|l|-1},
\label{borb}
\end{equation}
with ${\cal C}= 1/\sqrt{2^{|l|-1} \Gamma(|l|)}$ normalizing $|b\rangle$ to unity in
$[0,x_R]$.

From the theorem discussed above, it follows that the $\chi^A$ LLL component cannot have any 
zeros, and thus for $x_R=R/l_B >> 1$ it will develop into an edge state 
\begin{equation}
|e\rangle = -|\widetilde{\cal C}| e^{x^2/4} x^{|l|},
\label{eorb}
\end{equation}
due to the asymptotic behavior \cite{abrabook,note32}
\begin{equation}
M( 1-\varepsilon^2/2, |l|+1, x^2/2) \sim \exp(x^2/2) x_R^{-2|l|} 
\label{asbeh}
\end{equation}
for $x \sim x_R$. The normalization constant $\widetilde{\cal C}$ conforms to the requirement
that $\chi^A$ and $\chi^B$ [see Eq.\ (\ref{sp})] should have equal weights for the 
$\varepsilon$'s obtained via solution of Eq.\ (\ref{eigeq}); $\widetilde{\cal C}$ is 
determined from $\widetilde{\cal C}^2 \int_0^{x_R} e^{x^2/2} x^{2|l|+1} dx =1$.

Finally, for $R/l_B >> 1$, the following simplified expression for the LLL spinor can be
written:
\begin{equation}
\psi_l({\bf r})=
\frac{1}{\sqrt{2}}\left(
\begin{array}{c}
e^{il\phi} |e\rangle /\sqrt{2 \pi}\\
i e^{i(l+1)\phi} |b\rangle /\sqrt{2 \pi}
\end{array}
\right),
\label{finspi}
\end{equation}
where $|b\rangle$ and $|e\rangle$ are each normalized to unity in $[0,x_R]$. 
A numerical graphical illustration showing that the general expression for the spinor 
components in Eq.\ (\ref{solneg}) can be well approximated by Eq.\ (\ref{finspi}) is given in
Appendix B.  

\begin{figure}[t] 
\centering\includegraphics[width=7cm]{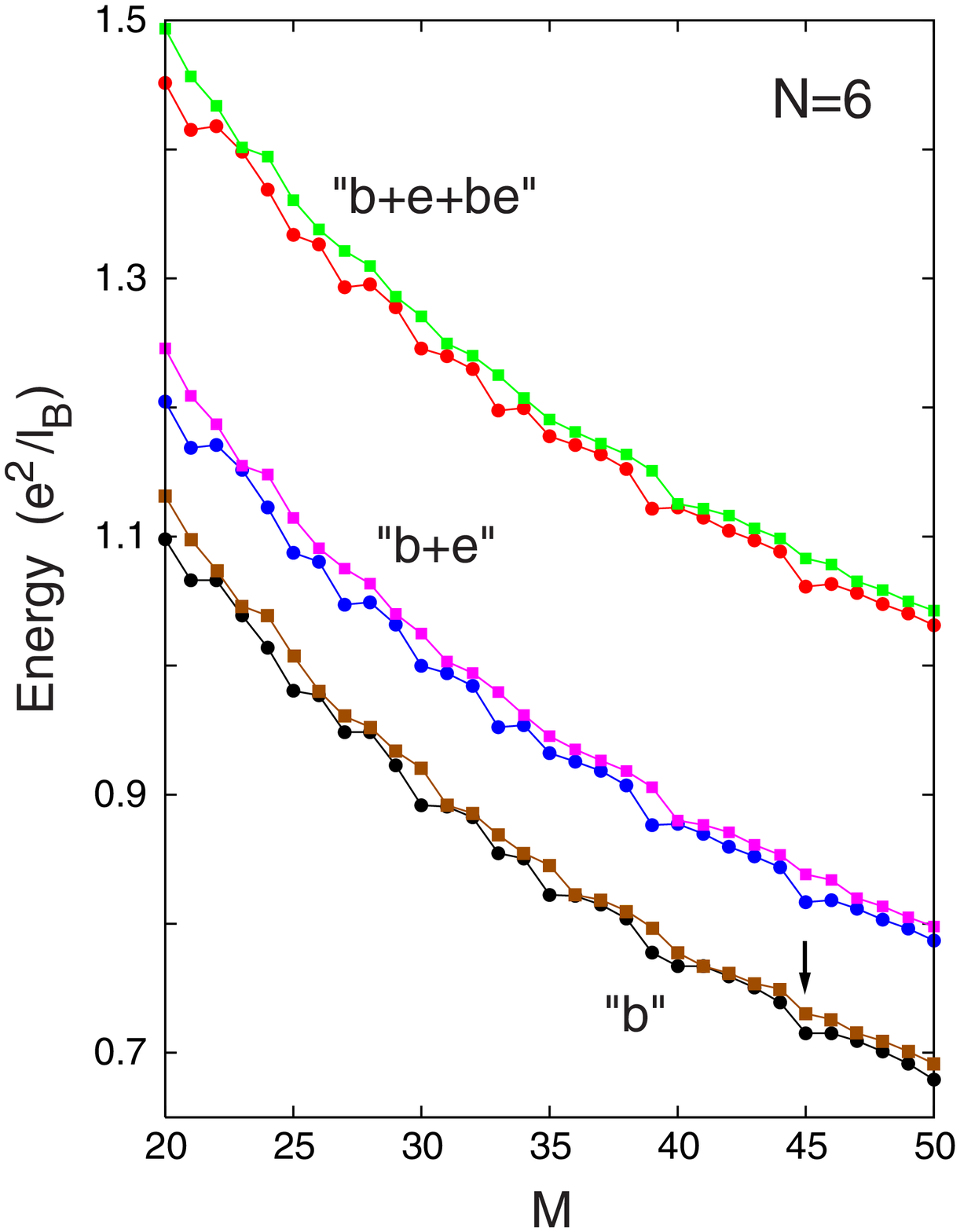}
\caption{
EXD energies for $N=6$ Dirac electrons. Energies for the lowest (yrast, solid dots) and 
first-excited states (solid squares) are plotted as a function of the total angular momentum 
$M=\sum_{i=1}^N m_i$ ($m_i \equiv |l_i|$). ``b'', ``e'', and ``be'' denote, respectively, the 
bulk, edge, and cross bulk-edge contributions to the total energies. The ``b'' contribution 
reflects a 50\% depletion of the bulk component due to the form of the Dirac-electron spinor in 
Eq.\ (\ref{finspi}). The arrow in the ``b'' curve points to the $\nu=1/3$ ($M=45$) excitation
energy gap (the difference between first-excited and yrast energies).
The radius of the sample was taken as $R=30 l_B$. $\kappa=1$ for suspended
graphene.
}  
\label{enen6}
\end{figure}

\begin{figure}[t] 
\centering\includegraphics[width=7cm]{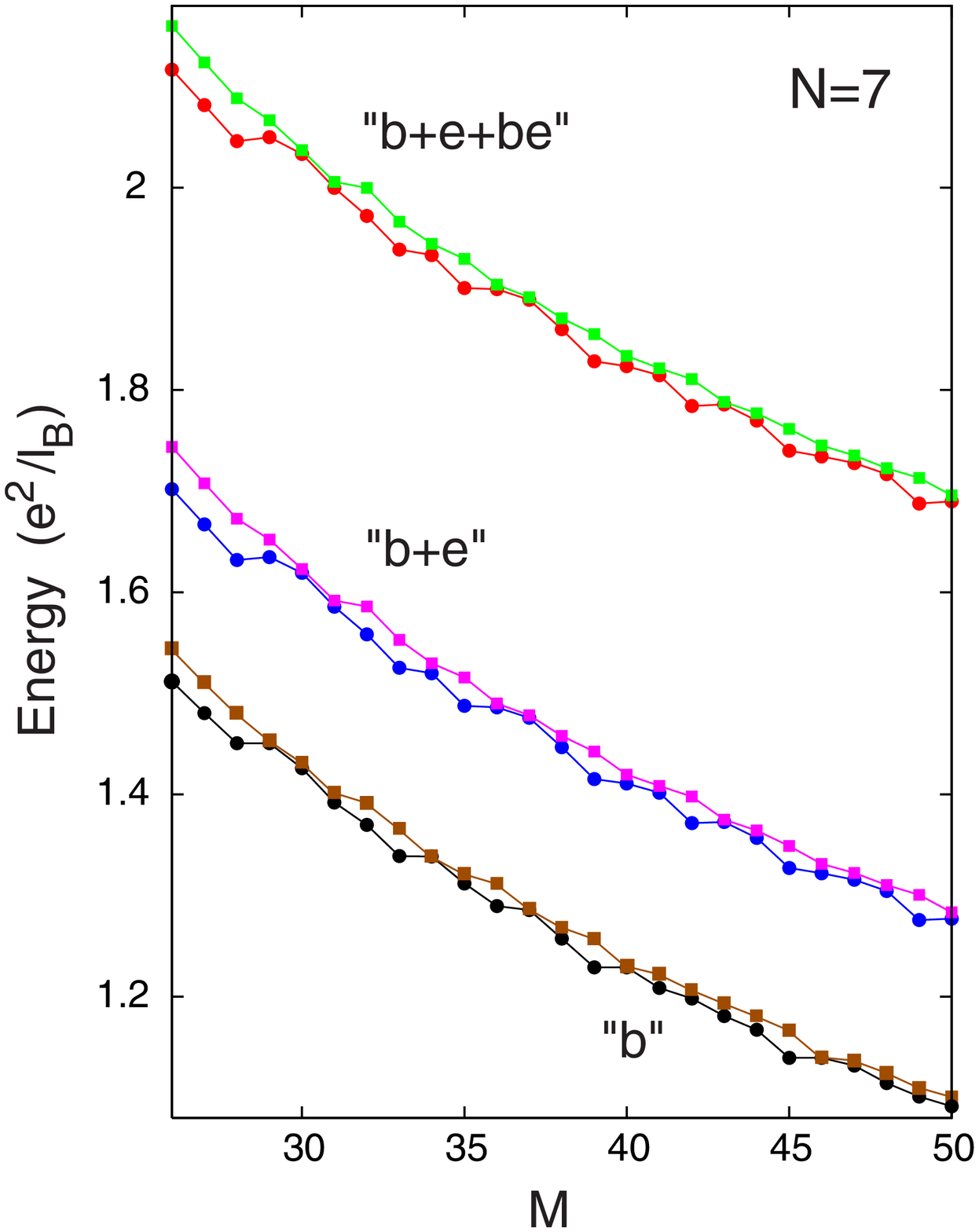}
\caption{
EXD energies for $N=7$ Dirac electrons. Energies for the lowest (yrast, solid dots) and 
first-excited states (solid squares) are plotted as a function of the total angular momentum 
$M=\sum_{i=1}^N m_i$ ($m_i \equiv |l_i|$). ``b'', ``e'', and ``be'' denote, respectively, the 
bulk, edge, and cross bulk-edge contributions to the total energies. The ``b'' contribution 
reflects a 50\% depletion of the bulk component due to the form of the Dirac-electron spinor in 
Eq.\ (\ref{finspi}). The radius of the sample was taken as $R=25 l_B$. $\kappa=1$ for suspended
graphene.
}  
\label{enen7}
\end{figure}

\section{Exact diagonalization of the many-body problem}
\label{secexd}

In the EXD method, \cite{yann07} the many-body wave function is given as a linear superposition
\begin{eqnarray}
\Phi^{\text{EXD}}_M
[{\bf r}_1,\ldots,{\bf r}_N]&=& \nonumber \\
&& \hspace{-3.8cm} \sum_{m_1<\ldots<m_N}^{m_1+\ldots+m_N=M}
{\cal V}(m_1,\ldots,m_N)
\Psi^{m_1,\ldots,m_N}[{\bf r}_1,\ldots,{\bf r}_N],
\label{wfexd}
\end{eqnarray}
where the Slater determinants $\Psi^{m_1,\ldots,m_N}$ are constructed out of the
quasidegenerate ($\varepsilon \sim 0$) LLL spinor orbitals $\psi_l({\bf r})$ [see Eq.\
(\ref{finspi})], where we define $m=|l|$; $l \leq -1$. We omit the 
Coulomb coupling between valleys, and also consider fully spin-polarized electrons.

The eigenvectors ${\cal V}$'s in Eq.\ (\ref{wfexd}) and corresponding many-body
energies $E^{\text{EXD}}$ are calculated through a direct matrix diagonalization of the
many-body hamiltonian
\begin{equation}
H=\sum_{i<j} V_C(i,j),
\label{mbh}
\end{equation}
where $V_C(i,j)=e^2/(\kappa |{\bf r}_i-{\bf r}_j|)$ is the two-body Coulomb interaction, within
the Hilbert space of Slater determinants \cite{yann07} $\Psi^{m_1,\ldots,m_N}$. 
Being constant (here $\approx 0$), the kinetic energy terms in the many-body hamiltonian
Eq.\ (\ref{mbh}) have been neglected. \cite{yann07,jainbook}

The EXD computations require an accurate evaluation of the two-body Coulomb matrix elements
\begin{equation}
\int\int d{\bf r}_1 d{\bf r}_2 \psi^*_{m_1}({\bf r}_1) \psi^*_{m_2}({\bf r}_2)
V_C(1,2) \psi_{m_3}({\bf r}_1) \psi_{m_4}({\bf r}_2), 
\label{cmesp}
\end{equation}
which expand to a sum of four similar integrals. Denoting $|\tilde{b}_m\rangle=
i e^{i(l+1)\phi}|b\rangle /\sqrt{2\pi}$ and
$|\tilde{e}_m\rangle=e^{il\phi}|e\rangle /\sqrt{2\pi}$, this expansion is written as
\begin{equation}
\frac{1}{4}( \langle \tilde{b}_1 \tilde{b}_2 | \tilde{b}_3 \tilde{b}_4 \rangle +
\langle \tilde{e}_1 \tilde{e}_2 | \tilde{e}_3 \tilde{e}_4 \rangle +
\langle \tilde{b}_1 \tilde{e}_2 | \tilde{b}_3 \tilde{e}_4 \rangle +
\langle \tilde{e}_1 \tilde{b}_2 | \tilde{e}_3 \tilde{b}_4 \rangle).
\label{cmeexp}
\end{equation}
Note that, due to the equal weights of the bulk-like and edge-like
components, a prefactor of 1/4 appears in front of each term in Eq.\ (\ref{cmeexp});
the consequences of this prefactor are discussed in Sec. \ref{secresene}.

\section{Results}
\label{secres}

\subsection{Energetics}
\label{secresene}

In Figs.\ \ref{enen6} and \ref{enen7}, we display total EXD energies for the yrast (i.e., the
lowest in energy) and first-excited states of spin-polarized electrons in a graphene finite 
sample with $N=6$ and $N=7$ Dirac electrons as a function of the total angular momentum $M$ 
(which is conserved in the EXD calculation). Moreover, the displayed energies correspond to: 
(i) only the bulk-type terms (``b''), (ii) combined bulk-type and edge-type terms (``b$+$e''), 
and (iii) with the inclusion of cross bulk-edge terms (``b$+$e$+$be'') in the two-body Coulomb 
matrix elements [see Eq.\ (\ref{cmeexp})]. Each of the terms (``b'', ``e'', and ``be'') is 
reduced by a factor of 1/4 as discussed above. For all cases, we note the appearance
of cusp states (states with enhanced stability and larger excitation gaps relative to 
their immediate neighborhood \cite{yann07,jainbook,laug83.2}) at the magic angular momenta 
\begin{equation}
M_m=M_0+kN \mbox{~~or~~} M_m=M_0+k(N-1), 
\label{mam}
\end{equation}
where $M_0=N(N-1)/2$, $k=0,1,2,3,\ldots$, and $N$ is the number of electrons. For the bulk 
component, the cusp states are interpreted \cite{jainbook,laug83.2} as a signature of the 
formation of correlated incompressible states that may exhibit certain correlated liquid 
characteristics and underlie the FQHE physics in the thermodynamic limit. In particular, the 
magic angular momenta are associated \cite{jainbook,laug83.2} with fractional fillings 
$\nu=M_0/M_m$. Note that the angular momenta of the Laughlin FQHE function 
\cite{laug83} correspond to Eq. (\ref{mam}) with $k=(N-1)p$ for the first case and $k=Np$ for 
the second case, with the associated fractions given by $\nu=1/(2p+1)$ ($p$ being a positive 
integer).

The EXD energies displayed in Fig.\ \ref{enen6} and \ref{enen7} [and additional calculated 
results for other sizes $N=3-5$ and $8$ (not shown)] reveal three prominent trends:

(I) The ``b$+$e'' energies are always higher than those calculated only with the bulk-type 
component (denoted as ``b''). More importantly (as checked for various sample sizes, e.g., 
$R/l_B=15, 25$ and 30), the difference between the ``b$+$e'' (i.e., including the edge 
contribution) and ``b'' EXD energies (for any 
$N$) approaches asymptotically for $M\rightarrow \infty$ the value $\Delta E_1(N)/4$;
$\Delta E_1(N)$ is the electrostatic energy of $N$ point charges localized on the 
perimeter at the vertices of a regular polygon, i.e.,
\begin{equation}
\Delta E_1(N)= 
\frac{N S_N}{4 x_R} \frac{e^2}{\kappa l_B},
\label{ecln}
\end{equation}
with $S_N= \sum_{j=2}^{N} \big(\sin[(j-1)\pi /N]\big)^{-1}$. The factor 1/4 is due to the half
weight of the electrons accumulating on the physical edge [recall the spinor
in Eq.\ (\ref{finspi})]. The result in Eq.\ (\ref{ecln}) (deduced from analysis of the EXD 
computed energies, see, e.g., Figs.\ \ref{enen6} and \ref{enen7}) suggests that a single-ring
Wigner molecule is formed at the edge; this is in agreement with the CPD analysis in Sec.
\ref{seccpds}. 

\begin{figure}[t]
\centering\includegraphics[width=7.5cm]{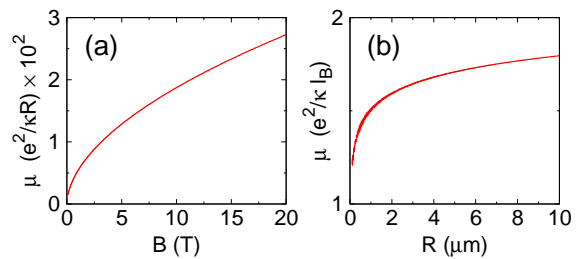}
\caption{
Variation of the insulating gap $\mu$. (a) As a function of $B$ when $R=1$ $\mu$m. 
(b) As a function of $R$ when $B=8$ T. Note the $\mu \propto B^{0.54}$ increase shown in (a)
and the effective saturation for large $R$ exhibited in (b). $\alpha=2$.
}
\label{engap}
\end{figure}

(II) The ``b$+$e$+$be'' energies are always higher than the ``b$+$e'' ones.
Moreover, the difference between the ``b$+$e$+$be'' and ``b$+$e'' EXD energies (for any 
$N$) approach asymptotically for $M\rightarrow \infty$ the value $\Delta E_2(N)/4$, where 
$\Delta E_2(N)$ is the electrostatic energy due to the repulsion between a point charge 
of strength $Ne$ located at the center and $Ne$ charges distributed at the perimeter of the 
sample,
\begin{equation}
\Delta E_2(N)= 
\frac{N (N-1)} {x_R} \frac{e^2}{\kappa l_B}.
\label{ecln2}
\end{equation}
The factor 1/4 is again due to half of the electrons being accumulated on the physical edge,
while the other half being located in the bulk component; the factor $N-1$ accounts for the
self-interaction correction. 

It is apparent that the EXD density of states associated with the ``b$+$e$+$be'' LLL spectrum 
will show (for a given $N$) the opening of an interaction-induced energy gap 
\begin{equation}
\Delta E(N) = \Delta E_1(N)/4 + \Delta E_2(N)/4
\label{eclnsum}
\end{equation}
compared to that associated with the 
bulk-only (``b'') spectrum. The emergence of this edge-induced, electrostatic energy gap 
is related to (see below) the experimentally observed \cite{andr09,kim09} insulating behavior 
at the charge neutrality point ($\nu=0$). 

\begin{figure}[t] 
\centering\includegraphics[width=4cm]{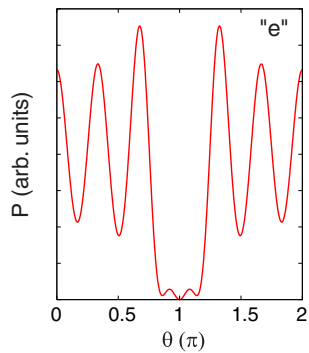}%
\caption{
CPD for the edge component portraying the formation of a single-ring RWM.
An illustrative yrast cusp state (with $N=6$ and $M=39$) in a 
graphene sample of radius $R/l_B=30$ is considered.
The azimuthal CPD at the physical edge of the sample is plotted, i.e., 
${\cal P}({\bf r}, {\bf r}_0)$ as a function of ${\bf r}=30(\cos\theta, \sin\theta)$.
The fixed point is located at ${\bf r}_0=30 (\cos\pi,\sin\pi)$. Note the five maxima
corresponding to the remaining five electrons, with the sixth one taken at ${\bf r}_0$.
${\bf r}$ and ${\bf r}_0$ are in units of $l_B$.
}  
\label{gdcpds}
\end{figure}

(III) Due to the fact that both the ``e'' and ``e$+$be'' contributions (expressed in units of 
$e^2/\kappa l_B$) decrease as a function of $R/l_B$ [see points (I) and (II) above, and the
factor $1/x_R=l_B/R$ in Eqs. (\ref{ecln}) and (\ref{ecln2})], it 
follows that in the thermodynamic limit the excitation gaps (which determine the strength of 
the FQHE) will approach those of the ``b'' contribution alone (which is independent of 
$R/l_B$). As a result, the FQHE excitation energy gaps for a graphene sample with zigzag edges
(and in particular for $\nu=1/3$) are 1/4 [see the prefactor of the bulk-only contribution in 
Eq.\ (\ref{cmeexp})] of those calculated \cite{jain06,chak06} using boundless-graphene 
modeling. For boundless graphene, available in the literature EXD calculations in a spherical 
geometry reported two values for the FQHE energy gap: the first value \cite{andr09,chak06} 
is $\sim$ 0.1 $e^2 /\kappa l_B$ for polarized electrons, while the second value
\cite{jain06} is 0.042 $e^2 /\kappa l_B$ when pseudoskyrmion effects are considered. 
\cite{note12} From our disk-geometry EXD calculations, we 
find (in units of $e^2/\kappa l_B$, and including the above 1/4 factor) an excitation energy 
gap for the ``b'' component of 0.0153 ($N=6$, $M_{1/3}=45$), 0.0160 ($N=7$, $M_{1/3}=63$), 
0.0145 ($N=8$, $M_{1/3}=84$), 0.0150 ($N=9$, $M_{1/3}=108$), 0.0149 ($N=10$, $M_{1/3}=135$), 
and 0.0147 ($N=11$, $M_{1/3}=165$) ($M_{1/3}$ gives the magic angular momenta corresponding to
$\nu=1/3$). These values extrapolate to an excitation energy gap of $0.0137$ $e^2/\kappa l_B$ 
for $N \rightarrow \infty$ (with a standard error of $\pm 0.0009$). Our result is in good 
agreement with the experimentally measured \cite{andr09,kim09} value 
(0.008 $e^2/\kappa l_B$), with a possible added effect of residual disorder.

To estimate the magnitude of the experimentally observable insulating gap, 
$\mu (\nu=0) =  \Delta E (N_{\text{max}}+1 ) - \Delta E (N_{\text{max}})$ [recall Eq.\
(\ref{eclnsum})], one needs to consider the maximum number, $N_{\text{max}}$, of RWM electrons 
on the edge of the graphene disk [see point (I) above]. A rough estimate is given by
$N_{\text{max}} \sim 2 \pi R/(2 \alpha l_B)$, where $R$ is the radius of the 
graphene sample and $\alpha$ is a fitting parameter. The variation of $\mu (\nu=0)$ with $B$
shown in Fig.\ \ref{engap}(a) (calculated for $R=1$ $\mu$m) is found to behave as
$B^\gamma$, $\gamma \approx 0.54$; the increase of the gap with $B$ is consistent
with recent experiments on suspended graphene. \cite{andr09,kim09} The evolution
of the insulating gap with the size of the system is found to essentially saturate 
for large $R$ [see Fig.\ \ref{engap}(b) calculated for $B=8$ T].

\subsection{Conditional probability distributions for the edge component}
\label{seccpds}

The calculated electron density is azimuthally uniform due to the conservation of the total 
angular momentum $M$ in the many-body EXD calculations. The spatial arrangement of the 
edge-type component of the $N$ correlated electrons can be revealed in the intrinsic frame of 
reference via the two-body correlation function \cite{yann07} ${\cal P}({\bf r}, {\bf r}_0)$, 
referred to as a conditional probability distribution (CPD). The CPD is defined as 
\begin{equation}
{\cal P}({\bf r}, {\bf r}_0) \propto
\langle \Phi^{\text{EXD}}_M | \sum_{i \neq j} 
\delta({\bf r}_i-{\bf r}) \delta({\bf r}_j-{\bf r}_0)
| \Phi^{\text{EXD}}_M \rangle.
\label{cpd}
\end{equation}
The CPDs are giving the 
probability of finding one electron at position ${\bf r}$  assuming that another one is 
fixed at the point ${\bf r}_0$. The CPD for the edge (``e'') component of the yrast cusp state
having $N=6$ and $M=39$ is displayed in Fig.\ \ref{gdcpds}. It is apparent that an RWM 
reflecting a single-ring arrangement is formed at the edge of the graphene sample.

\begin{figure}[t]
\centering\includegraphics[width=7cm]{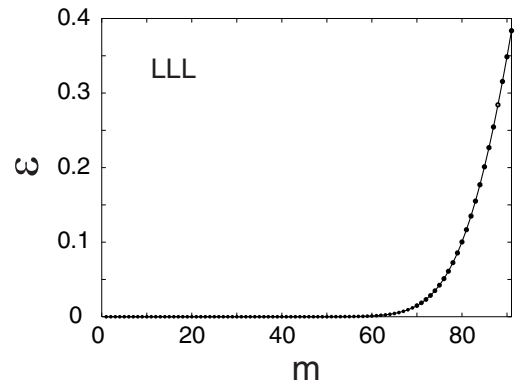}
\caption{
LLL reduced single-particle energies $\varepsilon$ as a function of the single-particle angular 
momentum $m=|l|$. Note the flat section of the curve that is followed by a rapid rise when the
$\chi^B$ bulk-type component reaches the edge of the graphene sample. The graphene sample
boundary is taken at $R=15 l_B$. $\varepsilon$ is dimensionless (see text).     
}
\label{eneLLL}
\end{figure}

\section{Summary}
\label{secsum}

We developed here a consistent picture of the many-body properties of interacting correlated 
Dirac electrons in graphene which is in correspondence with experimental findings pertaining 
to the incompressible FQHE states and insulating phase emerging at high magnetic fields. Key 
to the success of our theoretical model is the proper inclusion of the effect of the 
graphene-sample edge, treated here with the use of the zigzag boundary condition. This BC 
weights by a factor $1/\sqrt{2}$ the bulk-type wave functions in the two-component LLL Dirac 
spinor [Eq.\ (\ref{finspi})]; accounting for a 50\% depletion of the bulk electron density.
The other component (also weighted by $1/\sqrt{2}$) resides on the sample edge, leading to the 
emergence of an insulating phase. 

These findings point to the unique role of the graphene edge
in modulating the strength of the interelectron interactions governing the correlated 
many-body states in graphene, thus providing an impetus for future explorations, including 
controlled treatments of the graphene-sample boundaries.

\begin{acknowledgments}
We thank the referee for insightful suggestions. This work was supported by the US D.O.E. 
(Grant No. FG05-86ER45234)
\end{acknowledgments}

\begin{figure}[t]
\centering\includegraphics[width=8cm]{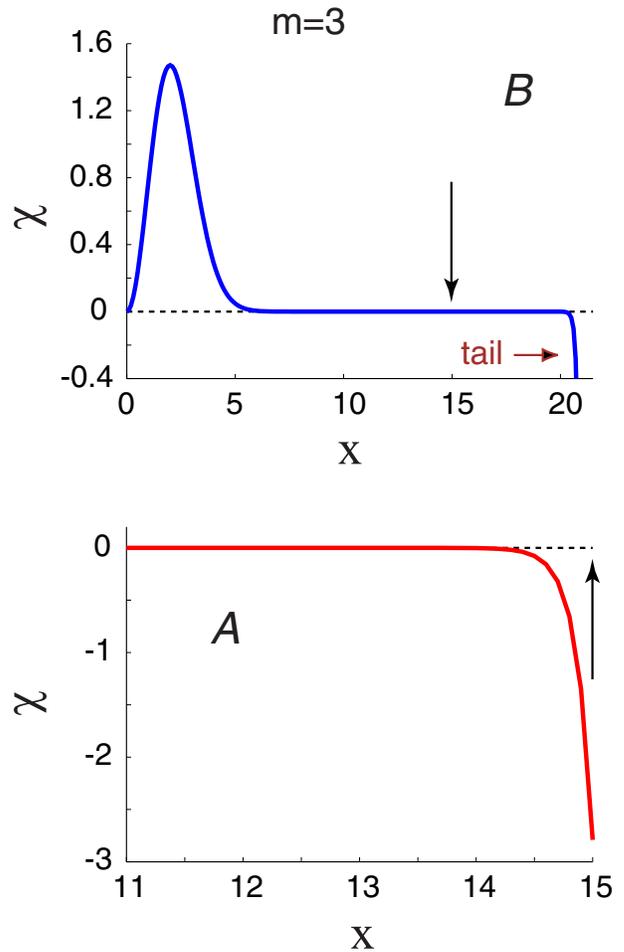}
\caption{
The two LLL orbital components $\chi^B(x)$ (marked as $B$, top frame) and $\chi^A(x)$ (marked 
as $A$, bottom frame) in Eq.\ (\ref{solneg}) for $m=3$ ($l=-3$) and $x_R=15$ (a mixed bulk-edge
DW spinor). The corresponding single-particle energy [see Eq.\ (\ref{eigeq}) and Fig.\ 
\ref{eneLLL}] is $\varepsilon=4.383 \times 10^{-22}$. The vertical arrows at $x=x_R=R/l_B=15$ 
mark the position of the physical edge of the graphene sample. Inside the graphene sample 
($x < x_R$), the $\chi^B$ component can be approximated by the bulk-type orbital in Eq.\ 
(\ref{borb}); outside the graphene sample ($x > x_R$), it develops an exponentially growing 
tail reflecting the fact that the full wave function in the entire range $0 \leq x < \infty$ 
is described via a confluent hypergeometric function exhibiting a zero at $x=x_R$. The
$\chi^A$ component is well represented by the edge orbital in Eq.\ (\ref{eorb}).
}
\label{becomp}
\end{figure}

\appendix

\section{Solutions of the Dirac-Weyl equation in polar coordinates}

The derivation of the solutions of the Dirac-Weyl Eq.\ (\ref{dweq}) involves two
subcases, i.e, for $\varepsilon \neq 0$ and $\varepsilon = 0$.

For $\varepsilon \neq 0$, the solutions of Eq.\ (\ref{dweq}) that are {\it regular at
the origin\/} ($x=0$) can be found through the substitution of the following form in the
DW equation,
\begin{equation}
\chi^A = C e^{-x^2/4} x^{|l|} f_l(x),
\label{regsub}
\end{equation}
which will specify $f_l(x)$.

For any $l$, and prior to invoking the zigzag boundary condition, the $\chi^A$ component is 
given by
\begin{eqnarray}
\chi^A(x) &=& C e^{-x^2/4} x^{|l|} \times \nonumber \\
&& M \left( \frac{|l|+l}{2}+1-\frac{\varepsilon^2}{2}, |l|+1, \frac{x^2}{2} \right).
\label{chia}
\end{eqnarray}
\noindent
In Eq.\ (\ref{chia}), $\varepsilon=E/(\hbar v_F/l_B)$ is the reduced Dirac-electron energy, 
with $v_F$ the Fermi velocity and $l_B=(\hbar c/eB)^{1/2}$ the magnetic
length. The general form of the $\chi^B$ component depends on whether $l$ is positive or
negative, i.e.,
\begin{equation}
\chi^B(x) = -C \frac{2|l|}{\varepsilon} e^{-x^2/4} x^{|l|-1}
M \left( -\frac{\varepsilon^2}{2}, |l|, \frac{x^2}{2} \right),
\label{chibn}
\end{equation}
for $l \leq -1$, and
\begin{eqnarray}
\chi^B(x) &=& C \frac{\varepsilon}{2(l+1)} e^{-x^2/4} x^{l+1} \times \nonumber \\
&& M \left( l+1-\frac{\varepsilon^2}{2}, l+2, \frac{x^2}{2} \right),
\label{chibp}
\end{eqnarray}
for $l \geq 0$.

For $\varepsilon = 0$, the two equations in the Dirac-Weyl system [Eq.\ (\ref{dweq})]
decouple. In this case, the regular at the origin ($x=0$) solutions are:

\begin{eqnarray}
\chi^A(x) &=& 0 \nonumber \\
\chi^B(x) &=& C^B x^{|l|-1} e^{-x^2/4},
\label{solnegezero}
\end{eqnarray}
for $l \leq -1$, and

\begin{eqnarray}
\chi^A(x) &=& C^A x^l e^{x^2/4} \nonumber \\
\chi^B(x) &=& 0.
\label{solposezero}
\end{eqnarray}
for $l \geq 0$.

The interacting-electron EXD energies corresponding to these ($\varepsilon = 0$) 
single-particle levels (taken as a basis) are higher than those resulting from the bulk-edge 
mixed states ($\varepsilon \neq 0$); consequently they correspond to excited states lying above
the yrast band.

\begin{figure}[t]
\centering\includegraphics[width=8cm]{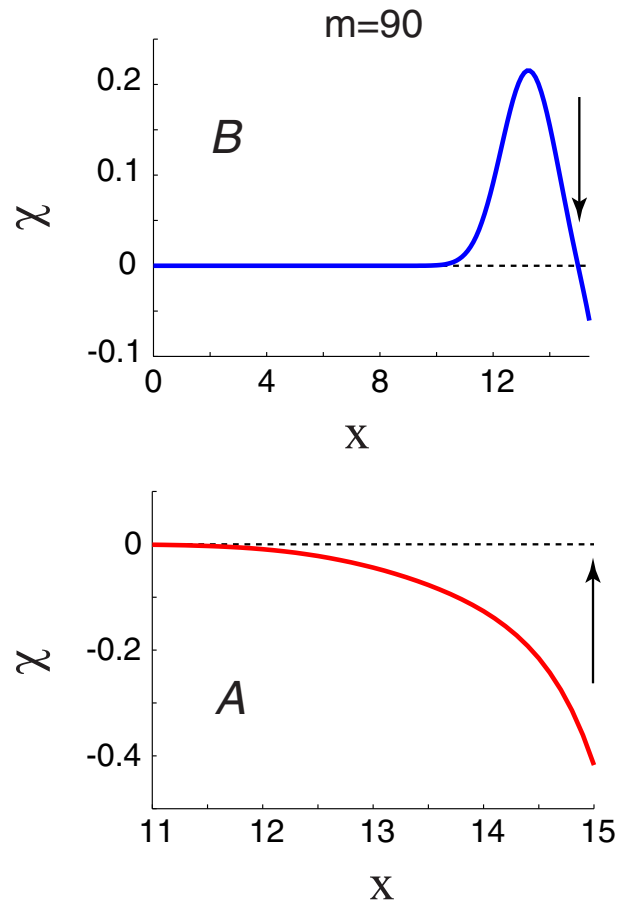}
\caption{
The two LLL orbital components $\chi^B(x)$ (marked as $B$, top frame) and $\chi^A(x)$ (marked 
as $A$, bottom frame) in Eq.\ (\ref{solneg}) for $m=90$ ($l=-90$) and $x_R=15$ (a 
``double-edge'' DW spinor, with both components representing edge-like orbitals). The 
corresponding single-particle energy [see Eq.\ (\ref{eigeq}) and Fig.\ \ref{eneLLL}] is 
$\varepsilon=0.349$. The vertical arrows at $x=x_R=R/l_B=15$ mark the position of the physical 
edge of the graphene sample. The $\chi^B$ component is similar to the nonrelativistic edge 
states familiar from the theory of the integer QHE in semiconductor 
heterostructures.\cite{halp82} The $\chi^A$ component is similar to the edge orbital in 
Eq.\ (\ref{eorb}).
}
\label{eecomp}
\end{figure}

\section{Two different types of edge states}

From early on, the concept of edge states has played a central role in the theory of the 
quantum Hall effect for nonrelativistic electrons.\cite{halp82} In addition to these
nonrelativistic edge states, the sublattice struture of graphene (relativistic, 
Dirac electrons) allows in high $B$ for other unique types of edge states with no analog in 
nonrelativistic systems; in particular, the mixed (bulk-edge) states described by the spinor in 
Eq.\ (\ref{finspi}). In this Appendix, we elaborate further on this point.  
  
To this effect, it is instructive first to investigate the behavior of the LLL energies 
$\varepsilon \neq 0$ obtained as solutions of the transcendental equation (\ref{eigeq}). 
For a graphene sample with radius $R=15 l_B$, the variation of the single-particle energies 
with $m=|l|$ is shown in Fig.\ \ref{eneLLL}. We observe a flat region where $\varepsilon$ is
vanishingly small ($\varepsilon \sim 0$) transforming for higher angular momenta to a steeply
rising branch. To gain insight into the nature of the single-particle levels in these two
regions, we plot the two components of the Dirac-electron spinor for two characteristic
angular momenta $m$.

In Fig.\ \ref{becomp} we show the case of $m=3$ ($l=-3$, with $\varepsilon=4.383 \times 
10^{-22}$), which lies deep inside the flat region in Fig.\ \ref{eneLLL}. Inside the graphene 
sample ($x < x_R$), the $\chi^B$ component can be approximated by the bulk-type orbital in Eq.\
(\ref{borb}); outside the graphene sample ($x > x_R$), it develops an exponentially growing
tail reflecting the fact that the full wave function in the expanded range $0 \leq x < \infty$
is described via a confluent hypergeometric function exhibiting a zero at $x=x_R$. The
$\chi^A$ component is well represented by the edge orbital in Eq.\ (\ref{eorb}), and as
aforementioned it does not have an analog for nonrelativistic LLL electrons.

We contrast this behavior by illustrating in Fig.\ \ref{eecomp} the spinor components
corresponding to $m=90$ ($\varepsilon=0.349$) which is characteristic of the steeply rising
branch in Fig.\ \ref{eneLLL}. The $\chi^B$ component is similar now to a nonrelativistic edge 
state familiar from the theory of the integer QHE in semiconductor 
heterostructures,\cite{halp82} while the $\chi^A$ component exhibits again a behavior
similar to the edge orbital in Eq.\ (\ref{eorb}). 

From the above we conclude that the origin of the rising energy of the single-particle states 
with larger $m$ (see Fig.\ \ref{eneLLL}) reflects the shift of the centroid of the $\chi^B$ 
component toward the boundary while (simultaneously) satisfying the vanishing of this component
on the sample boundary, thus disturbing the orbital shape (see Fig.\ \ref{eecomp}). Note that 
the position of the centroid depends on $m$ (compare Figs.\ \ref{becomp} and \ref{eecomp}) and 
that $\langle x^2 \rangle \approx 2(m+1)$ for $\sqrt{\langle x^2 \rangle} < x_R$. Working in 
the regime $R/l_B >> 1$, the states in the flat region of the energy curve in Fig.\ 
\ref{eneLLL} form an effective LLL manifold composed of mixed bulk-edge single-particle spinors 
[approximated by the Dirac-electron spinor in Eq.\ (\ref{finspi})].



\begin{thebibliography}{999}
\bibitem{geim04}
K.S. Novoselov {\it et al.\/},
Science {\bf 306}, 666 (2004).
\bibitem{geim05}
K.S. Novoselov {\it et al.\/},
Nature {\bf 438}, 197 (2005).
\bibitem{kim05}
Y. Zhang, Y.-W. Tan, H.L. Stormer, and Ph. Kim,
Nature {\bf 438}, 201 (2005).
\bibitem{jain06}
C. T\"{o}ke, P.E. Lammert, V.H. Crespi, and J.K. Jain,
Phys. Rev. B {\bf 74}, 235417 (2006).
\bibitem{chak06}
V.M. Apalkov and T. Chakraborty,
{Phys. Rev. Lett. {\bf 97}, 126801 (2006).
\bibitem{laug83}
R.B. Laughlin, 
Phys. Rev. Lett. {\bf 50}, 1395 (1983).
\bibitem{yann07}
C. Yannouleas and U. Landman,
Rep. Prog. Phys. {\bf 70}, 2067 (2007);
Phys. Rev. A {\bf 81}, 023609 (2010). 
\bibitem{jainbook}
J.K. Jain,
{\it Composite Fermions\/} (Cambridge University Press, 2007).
\bibitem{andr09}
X. Du, I. Skachko, F. Duerr, A. Luican, and E.Y. Andrei,
Nature {\bf 462}, 192 (2009).
\bibitem{kim09}
K.I. Bolotin, F. Ghahari, M.D. Shulman, H.L. Stormer, and Ph. Kim,
Nature {\bf 462}, 196 (2009).
\bibitem{bao10}
W. Bao {\it et al.\/}, arXiv:1005.0033v1 (2010).
\bibitem{geim09}
A.H. Castro Neto, F. Guinea, N.M.R. Peres, K.S. Novoselov, and A.K. Geim,
Rev. Mod. Phys. {\bf 81}, 109 (2009).
\bibitem{note56}
Using the high-precision capabilities of algebraic computer languages 
[see Ref.\ \onlinecite{mathbook}], we have checked numerically this equal-weight property. 
This property has been noted by L. Brey and H. A. Fertig [Phys. Rev. B\/} {\bf 73}, 
195408 (2006)] in the context of a study of edge states of noninteracting Dirac electrons in a 
graphene ribbon. The equal-weight property is related to the particle-hole symmetry of the
Dirac-Weyl equation (\ref{dweq}). Indeed, it follows immediately that if ($\chi^A$, $\chi^B$)
are solutions of Eq.\ (\ref{dweq}) with energy $\varepsilon$, then the pair 
($\chi^A$, $-\chi^B$) is also a solution with opposite energy $-\varepsilon$. For 
$\varepsilon \neq 0$, this corresponds to two different spinor solutions of Eq.\ (\ref{dweq})
which must be orthogonal; thus $\int (\chi^A)^2 rdr = \int (\chi^B)^2 rdr$.
\bibitem{peierls}
R. Peierls, 
{\it Surprises In Theoretical Physics\/} (Princeton University Press, Princeton, N.J., 1979),
Ch. 3.6.
\bibitem{abrabook}
M. Abramowitz and I.A. Stegun (Editors),
{\it Handbook Of Mathematical Functions\/},
(National Bureau of Standards, Washington, D.C., 1972).
\bibitem{mathbook}
See, e.g., S. Wolfram, {\it Mathematica: A System for Doing Mathematics by Computer\/} 
(Addison-Wesley, Reading, MA, 1991).
\bibitem{halp82}
B.I. Halperin, Phys. Rev. B {\bf 25}, 2185 (1982). 
\bibitem{gatt08}
W. Gautschi and C. Giordano,
Numer. Algor. {\bf 49}, 11 (2008), in particular Sect. 3.2.3. 
\bibitem{note32}
In particular formula 13.1.4 in Ref.\ \onlinecite{abrabook} and the
property that $M(0,b,z)=1$, $b>0$.
\bibitem{laug83.2}
R.B. Laughlin, 
Phys. Rev. B {\bf 27}, 3383 (1983).
\bibitem{note12}
For $\nu = 1/3$, the excitation gaps calculated from $N=4-7$ Dirac electrons in a spherical 
geometry (i.e., boundless graphene) in Ref.\ \onlinecite{jain06} are: (i) 0.042 (all energies 
are in units of $e^2/\kappa l_B$) [obtained from extrapolation of the $n=0$ (LLL) line in Fig.\ 
1(c) in the above cited paper], and (ii) 0.017 for the extrapolated value calculated for the 
$\nu=1/3$ filling of the $n=1$ Landau level [see the label ``$n=1$'' in Fig.\ 1(c)]. In both 
the $n=0$ and $n=1$ cases cited above, the gap involves composite-fermion pseudoskyrmions. The 
latter value (0.017) is the one cited in the experimental Ref.\ \onlinecite{kim09}. Without 
consideration of pseudoskyrmion effects, the excitation energy gap at the $\nu=1/3$ filling of 
the $n=0$ Landau level calculated in the spherical geometry in Ref.\ \onlinecite{chak06} [see 
Fig.\ 2(a) therein] for $N=8$ electrons is about 0.1 (see also Ref.\ \onlinecite{jainbook}, 
Table 6.5, p. 185). Ref.\ \onlinecite{andr09} cites this latter value (0.1) as 
the theoretically calculated one, noting that the measured $\nu=1/3$ excitation gap is 8\% of 
this value. The value 0.0137 obtained from extrapolation of our 
LLL $(n=0)$ results for $\nu =1/3$ is lower in an essential way from all the theoretical values
mentioned above, thus providing a potential signature for the edge-influenced depletion of the 
bulk Dirac-electron density.
\end{thebibliography}
\end{document}